\documentclass[twocolumn]{aastex631}

\usepackage{threeparttable}
\usepackage{subfigure}
\usepackage{longtable}
\begin{document}


\title{Resolving the 2024 Outburst of Magnetar 1E 1841--045 from its host Supernova Remnant with EP-FXT}

\author[0009-0005-2228-0618]{Yu-Cong Fu}
\affiliation{Institute for Frontiers in Astronomy and Astrophysics, Beijing Normal University, Beijing 102206, China}
\affiliation{School of Physics and Astronomy, Beijing Normal University, Beijing 100875, China}

\author[0000-0002-0633-5325]{Lin Lin}
\affiliation{Institute for Frontiers in Astronomy and Astrophysics, Beijing Normal University, Beijing 102206, China}
\affiliation{School of Physics and Astronomy, Beijing Normal University, Beijing 100875, China}

\author{Yu-Jia Zheng}
\affiliation{School of Physics and Technology, Nanjing Normal University, Nanjing 210023, Jiangsu, China}
\affiliation{State Key Laboratory of Particle Astrophysics, Institute of High Energy Physics, Chinese Academy of Sciences, Beijing 100049, China}

\author{Ming-Yu Ge}
\affiliation{State Key Laboratory of Particle Astrophysics, Institute of High Energy Physics, Chinese Academy of Sciences, Beijing 100049, China}
\affiliation{University of Chinese Academy of Sciences, Chinese Academy of Sciences, Beijing 100049, China}

\author[0009-0009-8477-8744]{Han-Long Peng}
\affiliation{School of Physics and Technology, Nanjing Normal University, Nanjing 210023, Jiangsu, China}

\author{Dong-Ming Li}
\affiliation{Institute for Frontiers in Astronomy and Astrophysics, Beijing Normal University, Beijing 102206, China}
\affiliation{School of Physics and Astronomy, Beijing Normal University, Beijing 100875, China}

\author{Francesco Coti Zelati}
\affiliation{Institute of Space Sciences (ICE, CSIC), Campus UAB, Carrer de Can Magrans s/n, Barcelona E-08193, Spain}
\affiliation{Institut d’Estudis Espacials de Catalunya (IEEC), Barcelona E-08034, Spain}

\author{Ersin G{\"o}{\v{g}}{\"u}{\c{s} }}
\affiliation{Sabancı University, Faculty of Engineering and Natural Sciences, İstanbul 34956, Turkey}

\author{Nanda Rea}
\affiliation{Institute of Space Sciences (ICE, CSIC), Campus UAB, Carrer de Can Magrans s/n, Barcelona E-08193, Spain}
\affiliation{Institut d’Estudis Espacials de Catalunya (IEEC), Barcelona E-08034, Spain}

\author[0000-0002-9725-2524]{Bing Zhang}
\affiliation{The Hong Kong Institute for Astronomy and Astrophysics, The University of Hong Kong, Pokfulam, Hong Kong, China}
\affiliation{Department of Physics, The University of Hong Kong, Pokfulam, Hong Kong, China}
\affiliation{Nevada Center for Astrophysics, University of Nevada, Las Vegas, USA}
\affiliation{Department of Physics and Astronomy, University of Nevada, Las Vegas, USA}

\author{Wei-Wei Zhu}
\affiliation{National Astronomical Observatories, Chinese Academy of Sciences, Beijing 100101, China}
\affiliation{Institute for Frontiers in Astronomy and Astrophysics, Beijing Normal University, Beijing 102206, China}

\author{Ke-Jia Lee}
\affiliation{Department of Astronomy, Peking University, Beijing 100871, China}
\affiliation{Kavli Institute for Astronomy and Astrophysics, Peking University, Beijing 100871, China}
\affiliation{National Astronomical Observatories, Chinese Academy of Sciences, Beijing 100101, China}

\author{Teruaki Enoto}
\affiliation{Department of Physics, Kyoto University, Kyoto 606-8502, Japan}

\author{Chryssa Kouveliotou}
\affiliation{Department of Physics, The George Washington University, Washington, DC 20052, USA}
\affiliation{Astronomy, Physics and Statistics Institute of Sciences (APSIS), The George Washington University, Washington, DC 20052, USA}

\correspondingauthor{Lin Lin}
\email{llin@bnu.edu.cn}




\begin{abstract}
The magnetar 1E 1841--045 exhibited a new active episode starting on August 20, 2024, marked by X-ray bursts and enhanced persistent emission.
Using data from the Einstein Probe (EP), we report on the timing and spectral results following the onset of this outburst.
The pulse profile displays a multi-peaked structure, with notable phase shifts in the secondary peak. 
Energy-resolved pulse profile analysis indicates a transition in the dominant peak of the pulse profile above 5.8 keV.
The 0.5--10 keV X-ray spectrum is well-modeled by a combined blackbody and power-law (BB+PL) model, showing a $\sim 20\%$ flux increase following the outburst.
Phase-resolved spectroscopy indicates a correlation between BB temperature and pulse profile intensity, along with spectral hardening at a specific pulse phase. 
The high spatial resolution of EP enables effective separation of the supernova remnant emission, which is crucial for measuring the intrinsic pulse emission of the source.
These findings underscore the intricate relationship between magnetar outbursts, pulse profile evolution, and spectral characteristics.
\end{abstract}



\section{Introduction} \label{sec:intro}
Magnetars are a unique class of isolated neutron stars (NSs) whose X-ray emissions, including persistent emission and X-ray bursts, are mainly powered by their extremely intense magnetic fields \citep[e.g.,][]{1992ApJ...392L...9D, 1993ApJ...408..194T, 1998Natur.393..235K}.
The magnetic fields are estimated from their spin periods $P$ (within a range of $\sim$ 1--12 s) and spin-down rates $\dot P$ (within a range of $\sim 10^{-13}-10^{-10}$s s$^{-1}$), mainly exceeding $\sim 10^{14}$ G \footnote{Refer to the online McGill SGR/AXP catalog, which contains the current information available on 30 magnetars, at: \href{http://www.physics.mcgill.ca/~pulsar/magnetar/main.html}{http://www.physics.mcgill.ca/\~{}pulsar/magnetar/main.html}} \citep{2014ApJS..212....6O}.
During the active period of magnetars, the brightening of persistent X-ray emission is frequently accompanied by short X-ray bursts, often along with the variation in spectral and timing properties \citep[e.g.,][]{2004ApJ...605..378W, 2017ApJ...851...17Y, 2025ApJ...980...99Y, 2026enap....3..205R}.

The soft X-ray emission observed in magnetars is predominantly attributed to NS surface emission, which is modified by the resonant scattering processes occurring within the magnetosphere \citep[e.g.,][]{2002ApJ...574..332T, 2006MNRAS.368..690L, 2010RAA....10..553T}. 
The persistent X-ray spectra below 10 keV typically require two components, commonly well-described by either an absorbed blackbody plus power-law (BB+PL) model or an absorbed double blackbody (BB+BB) model. 
These models are characterized by a typical BB temperature ($kT_{\rm BB}$) of $\sim 0.5$ keV and a photon index of $\sim 2-4$ in the BB+PL case; or by two distinct thermal components, with the cooler BB at $kT_{\rm cool}$ $\sim$ 0.3 keV and the hotter BB at $kT_{\rm hot}$ $\sim$ 0.7 keV, in the BB+BB case \citep{2014ApJS..212....6O, 2015SSRv..191..315M}.
During magnetar outbursts, the spectral evolution accompanying the decay in luminosity is often distinctive, characterized by phenomena such as spectral hardening and an increase in the inferred surface temperature \citep{2018MNRAS.474..961C}.
Additionally, temporal properties may also exhibit concurrent changes, including the occurrence of glitches/anti-glitches and the variation in the pulse profile \citep[e.g.,][]{2013Natur.497..591A, 2017ApJ...847...85Y, 2024RAA....24a5016G, 2025ApJ...980...99Y}.

1E 1841--045, initially discovered in 1985, is located at the center of the supernova remnant (SNR) Kes 73 (or G27.4+0.0, \citealt{1985ApJ...288..703K}). 
It was later identified as a pulsar in 1997 through the detection of its spin period \citep{1997ApJ...486L.129V}.
The spin period of 1E 1841--045 was calculated as $P\sim 11.8$ s, with a spin-down rate of $\dot P\sim 4 \times10^{-11}\rm s\,s^{-1}$, which implies a dipolar magnetic field strength of $ B\sim 7 \times10^{14}\rm\,G$ \citep{1997ApJ...486L.129V, 1999ApJ...522L..49G}.
The source is located at a distance initially estimated as $8.5^{+1.3}_{-1.0}$ kpc from the H I/II absorption and emission of Kes 73 \citep{2008ApJ...677..292T}, which has been updated to $5.8\pm0.3$ kpc from the re-analysis of H I absorption features \citep{2018AJ....155..204R}.
Over several decades, 1E 1841--045 has experienced occasional X-ray bursts, but its persistent soft X-ray emission has maintained a nearly constant luminosity \citep[e.g.,][]{1999ApJ...522L..49G, 2010ApJ...719..351Z, 2011ApJ...740L..16L, 2015ApJ...807...93A}.
Additionally, persistent hard X-ray emission in the range of 15 to 200 keV has been detected, characterized by a PL photon index of 1.3 \citep[e.g.,][]{2004ApJ...613.1173K, 2013ApJ...779..163A, 2015ApJ...807...93A}.
  
Deep radio observations of the magnetar 1E 1841--045 were conducted in previous studies.
The Parkes radio telescope (1.4 GHz, \citealt{2006MNRAS.372..410B}), Green Bank Telescope (GBT, 1950 MHz; \citealt{2012ApJ...744...97L}), Five-hundred-meter Aperture Spherical radio Telescope (FAST, 1250 MHz; \citealt{2025ApJ...979..122B}), MeerKAT (L-band and S-band), and Effelsberg radio telescopes (1.3--6.0 GHz; \citealt{2025arXiv250220079Y}) have yielded no detections of pulsed or single-pulse radio emission. 
The consistent lack of detections, despite using multiple sensitive telescopes across a range of frequencies, strongly suggests the source is radio quiet.

On 2024 August 20, 1E 1841--045 entered a new active period, triggering several instruments with a series of short bursts. 
Swift/BAT detected four bursts within the 15--350 keV energy range on August 20 and 21 \citep{2024GCN.37211....1D, 2024GCN.37222....1D}. 
Fermi/GBM also recorded eight bursts from this source between August 20 and 22, three of which were temporally coincident with Swift's detections \citep{2024GCN.37234....1R}. 
Additionally, burst activity was reported by SVOM/GRM \citep{2024GCN.37297....1S, 2024GCN.38192....1S}, GECAM \citep{2024GCN.37240....1Z}, INTEGRAL \citep{2024GCN.38148....1M}, NICER \citep{2024ATel16789....1N}, and Insight-HXMT \citep{2024GCN.37606....1C}. 
Notably, NICER observations revealed a $\sim 25\%$ increase in the persistent emission flux in the 0.5--10 keV range, along with a change in the structure of the pulse profile compared to the pre-burst (August 3) level \citep{2024ATel16789....1N, 2024ATel16802....1Y}.

During its 2024 active period, IXPE observations of 1E 1841--045 revealed a highly polarized X-ray emission, whose polarization degree (PD) changed from $\sim 15\%$ at 2--3 keV to $\sim 55\%$ at 6--8 keV \citep{2025ApJ...985L..34R, 2025ApJ...985L..35S}.
The 2--79 keV broadband spectrum was well-modeled by a combination of a BB and two PL components, with the hard PL component suggesting a synchrotron/curvature origin.
The soft PL component was thought to be the emission from a Comptonized corona in the inner magnetosphere \citep{2025ApJ...985L..35S}.
Following a spin-up glitch detection, NICER and NuSTAR observed a narrow peak in the pulse profile gradually moving closer to the main peak \citep{2025arXiv250220079Y}.

In this study, we report the results of Target of Opportunity (ToO) multi-wavelength observations of 1E 1841--045, carried out with the Einstein Probe and FAST.
The details of the observations and data reduction are provided in Section \ref{sec:Observations and Data Reduction}, the data analysis and results are described in Section \ref{sec:ANALYSIS AND RESULTS}, and the discussion and conclusions are given in Section \ref{sec:Discussion}.

\section{Observations and Data Reduction}
\label{sec:Observations and Data Reduction}
\subsection{X-ray observations}
The Einstein Probe (EP), launched on 2024 January 9, is a mission led by the Chinese Academy of Sciences \citep{2022hxga.book...86Y, 2025arXiv250107362Y}.
Focused on time-domain high-energy astrophysics, EP is equipped with two instruments: the Wide-field X-ray Telescope (WXT) and the Follow-up X-ray Telescope (FXT).
EP/WXT\footnote{\href{https://ep.bao.ac.cn/ep/cms/article/view?id=39}{https://ep.bao.ac.cn/ep/cms/article/view?id=39}}, a soft X-ray focusing telescope (0.5--4 keV), employs an exceptionally wide instantaneous field of view (FOV), covering 3600 square degrees, combined with a moderate spatial resolution of $\sim 5$ arcmin (Full Width at Half Maximum) and a time resolution of $\sim 50$ ms \citep{2024ApJ...975L..27Y, 2025SCPMA..6819511Z, 2025NatAs.tmp...21L}.
EP/FXT\footnote{\href{http://epfxt.ihep.ac.cn/about}{http://epfxt.ihep.ac.cn/about}}, a Wolter-I telescope operating in the 0.5--10 keV energy range, employs a narrow field of view with a diameter of 60 arcmin and a source localization error of 5--15 arcsec \citep{2020SPIE11444E..5BC}.
Consisting of two co-aligned identical units, FXT-A and FXT-B, FXT supports three scientific observation modes: Full-Frame Mode (FF, time resolution of 50 ms), Partial-Window Mode (PW, time resolution of 2.2 ms), and Timing Mode (TM, time resolution of 23.68 $\rm \mu s$).
Its primary role is to perform rapid follow-up observations (within 5 minutes) of the sources triggered by WXT, and will also observe other interesting targets during the all sky survey.

\begin{figure*}
    \centering  
    \includegraphics[width=\textwidth]{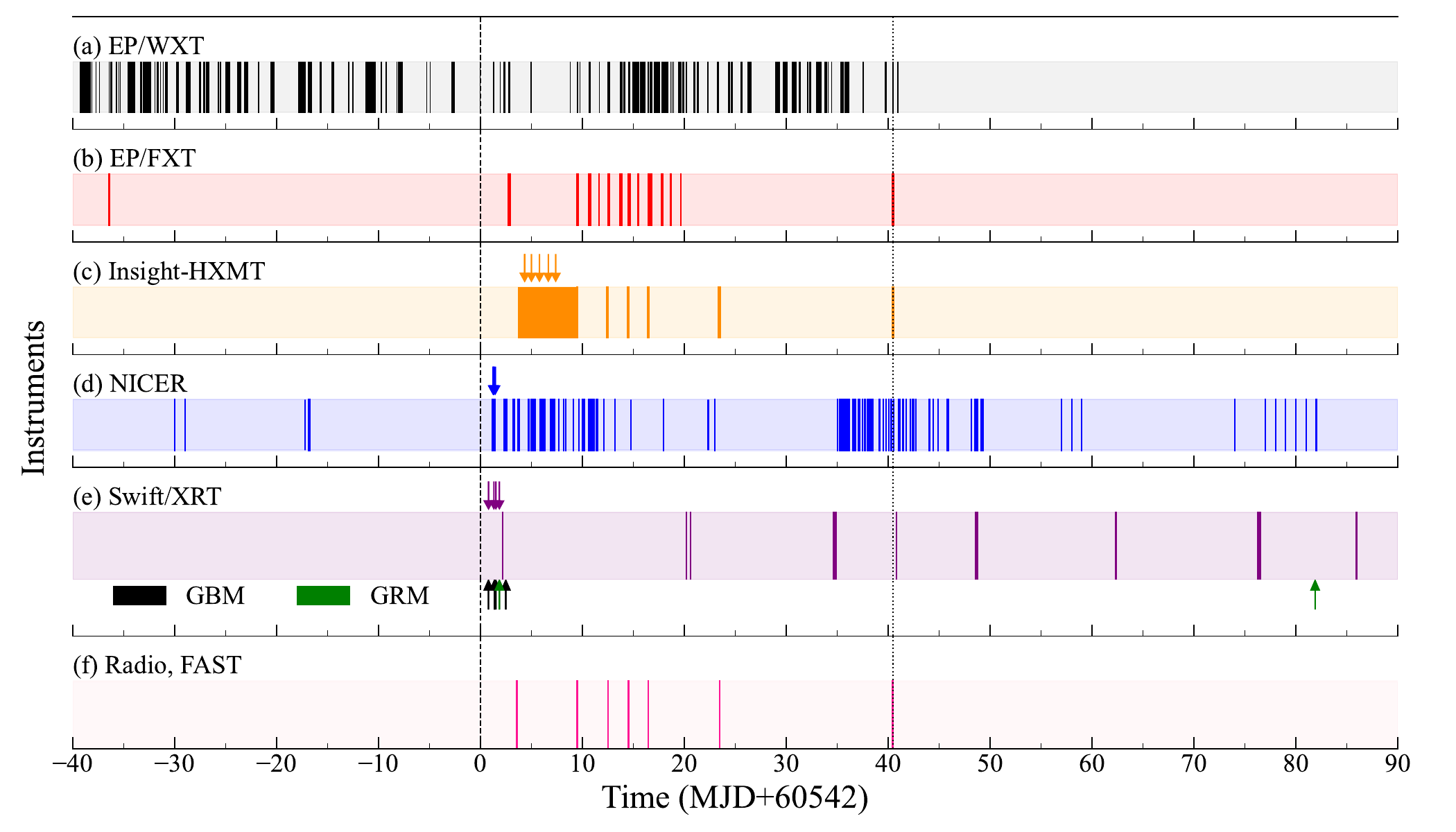}
    \caption{Observation timelines for different instruments.
    The dark vertical bars in each panel indicate the observation epochs. 
    The arrows mark the time of the short bursts reported in the references. 
    In panel (e), the purple arrows correspond to bursts detected by Swift/BAT, while the black and green arrows represent the bursts detected by Fermi/GBM and SVOM/GRM, respectively. 
    The dashed and dotted vertical lines represent the onset of the outburst and the epoch of a joint FAST and EP observation, respectively.}
    \label{fig:lcregion}
\end{figure*}

Following the detection of the burst activity, FXT conducted the ToO observations of 1E 1841--045 from 2024 August 22 to September 29, accumulating a total net exposure time of $\sim 85$ ks.
Additionally, FXT also observed the source during its Cycle-1 proposal before bursts (on 2024 July 14) with a net exposure time of $\sim 900$ s.
The observational details of FXT are summarized in Table \ref{tab:spec_table}.
During the all-sky monitoring, WXT covered the sky region containing the source and identified it in a total of 206 observations from 2024 May 17 to September 29.
The WXT observed flux\footnote{\href{https://ep.bao.ac.cn/ep/data_center/wxt_observation_data}{https://ep.bao.ac.cn/ep/data\_center/wxt\_observation\_data}} within the 0.5--4 keV range exhibits fluctuations ranging from 0.6 to 4.5, having an average value of 1.88 and a standard deviation of 0.61 (in units of $10^{-11}\,\rm erg \, s^{-1}\, cm^{-2}$).
Subsequently, 1E 1841--045 became invisible due to a solar aspect angle smaller than $94.5^\circ$.
The time intervals of WXT and FXT observations are illustrated in Figure \ref{fig:lcregion}.

\begin{longtable*}{lcccccccc}
    \caption{EP/FXT observations of 1E 1841--045.\label{tab:spec_table}}\\
	
    \hline  
    \hline 
	Obs. ID & Time &  Date  & Exp.$^{\rm a}$  & Mode & Rate$^{\rm b}$& Flux$^{\rm c}$&kT$_{\rm BB}$&$\Gamma^{\rm}$\\
    & (MJD)& (UTC) & (s) & &(cts/s)&($10^{-11}$)&(keV)&\\
    
    \hline
	11908432129 &	  60505.57   & 2024-7-14 & 936  &FF & $ 1.05 \pm 0.04 $ & $4.94 \pm 0.17$ & $0.47\pm0.03$ & $1.90^d$ \\
    \hline
    06800000053	&	60544.74 	& 2024-8-22 & 9009 &FF &$ 1.25 \pm 0.01 $ &$6.36\pm 0.06$&$0.46\pm0.02$&$1.99^{+0.10}_{-0.11}$\\
    10202120449	&	60551.49 	& 2024-8-29 & 2752 &FF &$ 1.21 \pm 0.02 $ &$5.88\pm 0.11$&$0.45\pm0.02$&$1.85^{+0.21}_{-0.25}$\\
    06800000059	&	60552.62 	& 2024-8-30 & 8967 &PW &$ 1.19 \pm 0.01 $ &$5.67\pm 0.06$&$0.48\pm0.01$&$1.80^{+0.12}_{-0.13}$\\
    06800000060	&	60553.62 	& 2024-8-31 & 1402 &PW &$ 1.14 \pm 0.03 $ &$5.47\pm 0.15$&$0.52\pm0.05$&$1.87^{+0.28}_{-0.48}$\\
    06800000061	&	60554.49 	& 2024-9-01 & 8958 &PW &$ 1.19 \pm 0.01 $ &$5.89\pm 0.06$&$0.50\pm0.02$&$1.99^{+0.09}_{-0.11}$\\
    06800000063	&	60555.67 	& 2024-9-02 & 8952 &PW &$ 1.20 \pm 0.01 $ &$5.73\pm 0.06$&$0.47\pm0.01$&$1.78^{+0.14}_{-0.16}$\\
    06800000068	&	60556.50 	& 2024-9-03 & 8951 &PW &$ 1.19 \pm 0.01 $ &$5.58\pm 0.06$&$0.47\pm0.01$&$1.60^{+0.14}_{-0.16}$\\
    06800000067	&	60557.43 	& 2024-9-04 & 5965 &PW &$ 1.16 \pm 0.01 $ &$5.30\pm 0.07$&$0.48\pm0.01$&$1.34^{+0.25}_{-0.29}$\\
    06800000071	&	60558.43 	& 2024-9-05 & 2982 &PW &$ 1.20 \pm 0.02 $ &$6.04\pm 0.06$&$0.51\pm0.02$&$2.06^{+0.09}_{-0.10}$\\
    06800000069	&	60558.63 	& 2024-9-05 & 8949 &PW &$ 1.20 \pm 0.01 $ &$6.26\pm 0.11$&$0.50\pm0.03$&$2.15^{+0.14}_{-0.17}$\\
    06800000070	&	60559.77 	& 2024-9-06 & 5966 &PW &$ 1.19 \pm 0.01 $ &$6.27\pm 0.08$&$0.51\pm0.03$&$2.13^{+0.09}_{-0.11}$\\
    06800000072	&	60560.64 	& 2024-9-07 & 2983 &PW &$ 1.22 \pm 0.02 $ &$5.93\pm 0.11$&$0.46\pm0.02$&$1.80^{+0.22}_{-0.27}$\\
    06800000076	&	60561.64 	& 2024-9-08 & 2984 &PW &$ 1.22 \pm 0.02 $ &$5.59\pm 0.10$&$0.46\pm0.01$&$1.49^{+0.24}_{-0.28}$\\
    08500000151	&	60582.42 	& 2024-9-29 & 5799 &PW &$ 1.18 \pm 0.02 $ &$6.34\pm 0.08$&$0.48\pm0.02$&$2.24^{+0.09}_{-0.11}$\\
    \hline
	
    \multicolumn{9}{l}{NOTES--}\\
    \multicolumn{9}{l}{$^{\rm a}$ Net exposure time (s) for each observation.}\\
    \multicolumn{9}{l}{$^{\rm b}$ The 0.5--10 keV net count rate of FXT-A and FXT-B.}\\
    \multicolumn{9}{l}{$^{\rm c}$ The 0.5--10 keV unabsorbed flux in units of $\rm 10^{-11} erg \, s^{-1} \, cm^{-2}$.}\\
    \multicolumn{9}{l}{$^{\rm d}$ The fixed parameter.}\\
\end{longtable*}

\begin{figure*}
    \centering  
    \includegraphics[width=0.8\textwidth]{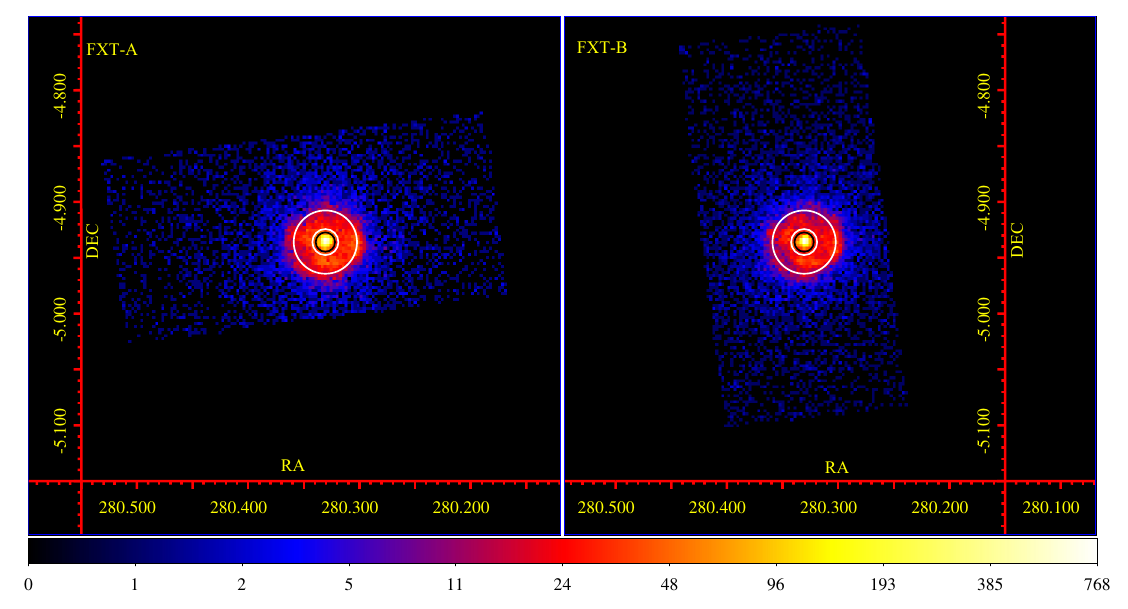}
    \caption{Logarithmic scale images of 1E 1841--045 captured by FXT-A (left) and FXT-B (right) of Obs. ID 06800000059 (MJD 60552.62). The blue rectangular areas indicate the FXT fields of view in PW mode. A black circle with a 30-arcsecond radius displays the central source. A white annulus with a 40-arcsecond inner radius and a 100-arcsecond outer radius displays the surrounding SNR. The color bar shows the counts per pixel.}
    \label{fig:image}
\end{figure*}

We processed the WXT observations using \texttt{wxtpipeline}, a dedicated analysis toolchain within the WXT Data Analysis Software (WXTDAS\footnote{\href{https://ep.bao.ac.cn/ep/cms/article/view?id=182}{https://ep.bao.ac.cn/ep/cms/article/view?id=182}}) and the calibration database (CALDB), both developed by the EP Science Center (EPSC). 
Applying the default calibration and screening procedures of \texttt{wxtpipeline}, we extracted the source light curves and spectra from a circular region with a radius of $300\farcs0$ to obtain as many source photons as possible. 
The FXT observations were processed with the \texttt{fxtchain} tool, which is a part of the FXT Data Analysis Software (FXTDAS\footnote{\href{http://epfxt.ihep.ac.cn/analysis}{http://epfxt.ihep.ac.cn/analysis}}, v1.10) developed by the EPSC.
Using the CALDB v1.10 and the default parameters, we extracted the clean event files, Good Time Interval (GTI) files, and image files.
As shown in Figure \ref{fig:image}, the source region was defined as a circular area with a radius of $30\farcs0$.
The background region, which is located within the SNR, was selected from an annulus with the inner and outer radii of $40\farcs0$ and $100\farcs0$, respectively.
Finally, the light curves and spectra were extracted using XSELECT v2.5b.

The events of the source region in the energy range of 0.5--4 keV (WXT) and 0.5--10 keV (FXT) after the barycenter correction were used for timing analysis.
We used the epoch-folding method \citep{1987A&A...180..275L} to determine the initial period of the NS \citep[e.g.,][]{2020ApJ...902L...2B, 2023MNRAS.521..893F, 2025ApJ...980...99Y}, and used the TEMPO2 \citep{2006MNRAS.372.1549E} v2024.12.1 to update the more accurate ephemeris of the source.
Times of arrival (ToAs) were derived through $Z^2$ searching, with the minimum phase in each profile serving as the ToA for the corresponding observation \citep{2012ApJS..199...32G, 2019NatAs...3.1122G, 2020ApJ...904L..21Y, 2026ApJ...999...85P}.
For spectral analysis, we performed fitting using XSPEC \citep{1996ASPC..101...17A} v12.14.1, as part of HEASoft v6.34, which allowed us to model and analyze the spectra comprehensively.

NICER data that partly overlap in time with the FXT observations are used to supplement the FXT timing dataset.
Standard processing, including the creation of cleaned and calibrated event files, is performed using the \texttt{nicerl2} tool within NICERDAS v12.
Further details and results of the NICER data processing are provided by \citet{2025arXiv250220079Y}.

\subsection{Radio observations}
The radio observations were carried out with the highly sensitive FAST telescope \citep{2020RAA....20...64J}. 
Following the X-ray activity, we conducted coordinated radio and X-ray observations of 1E 1841--045 using FAST and EP/FXT, respectively (see Figure \ref{fig:lcregion}). 
From August to September 2024, we obtained 7 FAST observations totaling 5.14 hours, with individual sessions typically lasting 30--60 minutes, as shown in Table \ref{tab:FAST}. 
These observations covered the 1--1.5 GHz band with high sensitivity, utilizing a setup of 4096 channels (0.122 MHz per channel) and a time resolution of 49.152 $\rm \mu$s.

\begin{deluxetable}{ccc}
\tablenum{2}
\tablecaption{FAST observations of 1E 1841--045\label{tab:FAST}}
\tablewidth{0pt}
\tablehead{
    \colhead{Observation Date} & \colhead{Start Time} & \colhead{Exposure} \\
    \colhead{(YYYYMMDD)} & \colhead{(MJD)} & \colhead{(minutes)}
}
\startdata
20240823 & 60545.553 & 38 \\ 
20240829 & 60551.492 & 60 \\ 
20240901 & 60554.487 & 30 \\ 
20240903 & 60556.517 & 60 \\ 
20240905 & 60558.458 & 30 \\ 
20240912 & 60565.458 & 30 \\ 
20240929 & 60582.444 & 60 \\ 
\enddata
\end{deluxetable}

\section{Results} \label{sec:ANALYSIS AND RESULTS}
\subsection{The joint observations and EP burst search }\label{sec:burstsearch}
After the onset of the 2024 August 20 active episode, multiple telescopes conducted follow-up observations.
The observation timelines for EP, Insight-HXMT, NICER, Swift/XRT, and FAST are summarized in Figure \ref{fig:lcregion}.
A notable coordinated multi-wavelength observation involving FAST and EP was performed simultaneously around MJD 60582.5.

We apply a Poissonian procedure to search for potential X-ray bursts for EP \citep[e.g.,][]{2004ApJ...607..959G}.
The initial light curves from EP/WXT are extracted with time resolutions of 0.05, 0.1, 0.2, and 0.5 seconds. 
No significant signals exceeding the $3 \sigma$ threshold are detected.
Additionally, no X-ray bursts were identified in the blind search of the EP/FXT data, with the significance of any candidate event failing to exceed $3 \sigma$ threshold above the persistent emission.
Compared with the reported bursts \citep{2024GCN.37222....1D, 2024GCN.37234....1R, 2024GCN.37297....1S, 2024GCN.38192....1S, 2024ATel16789....1N, 2024GCN.37606....1C} shown in Figure \ref{fig:lcregion}, the EP observations missed the most active epoch of bursts.

\subsection{Timing analysis}\label{sec:Results-1}
The statistical sample of the WXT events is insufficient to yield significant timing results.
In contrast, the periods detected by FXT exhibit high significance, with the optimal spin frequency $f$ and its derivative $\dot f$ determined to be 0.084699549(8) Hz and $-2.49(6)\times10^{-13}\, \rm Hz\, s^{-1}$, respectively, at a reference epoch of MJD 60549.
The ephemeris derived from FXT data is consistent with the results obtained from NICER data \citep{2025arXiv250220079Y}.
To improve statistical analysis, we combine the FXT and NICER observations during the same time interval.
As detailed in Table \ref{tab:tabletiming}, the best-fit to the FXT and NICER data spanning  MJD 60505 to MJD 60584 yields spin frequency $f$ and its derivative $\dot f$ are 0.084699564(6) Hz and $-3.04(5)\times10^{-13}\, \rm Hz\, s^{-1}$, respectively.
The resulting fit yields a reduced chi-square ($\chi^2$/dof) of 43.54/34 ($\sim$ 1.28), which is an improvement over the value of 24.24/17 ($\sim$1.43) from the FXT data alone.

\begin{deluxetable}{lcc}
\tablenum{3}
\tablecaption{Best fits spin parameters of 1E 1841--045 from EP/FXT and NICER data.\label{tab:tabletiming}}
\tablewidth{0pt}
\tablehead{
\colhead{Parameter} & \colhead{FXT and NICER }& \colhead{FXT}
}
\startdata
R.A. (J2000)&\multicolumn{2}{c}{18:41:19.6}\\ 
Decl. (J2000)&\multicolumn{2}{c}{$-$04:56:07.4}\\
Start (MJD)     &\multicolumn{2}{c}{60505} \\  
Finish (MJD)    &\multicolumn{2}{c}{60584}\\
Reference epoch (MJD)&\multicolumn{2}{c}{60549}\\ 
Spin frequency $f$ (Hz)& $0.084699564(6)$ & $0.084699549(8)$\\
Spin-down rate$\dot f\ (10^{-13}\, \rm Hz\, s^{-1})$&$-3.04(5)$ &$-2.49(6)$\\
\hline
$\chi^2/\rm dof$&43.54/34 &24.24/17\\
\enddata

\end{deluxetable}

\begin{figure}
    \centering  
    \includegraphics[width=\columnwidth]{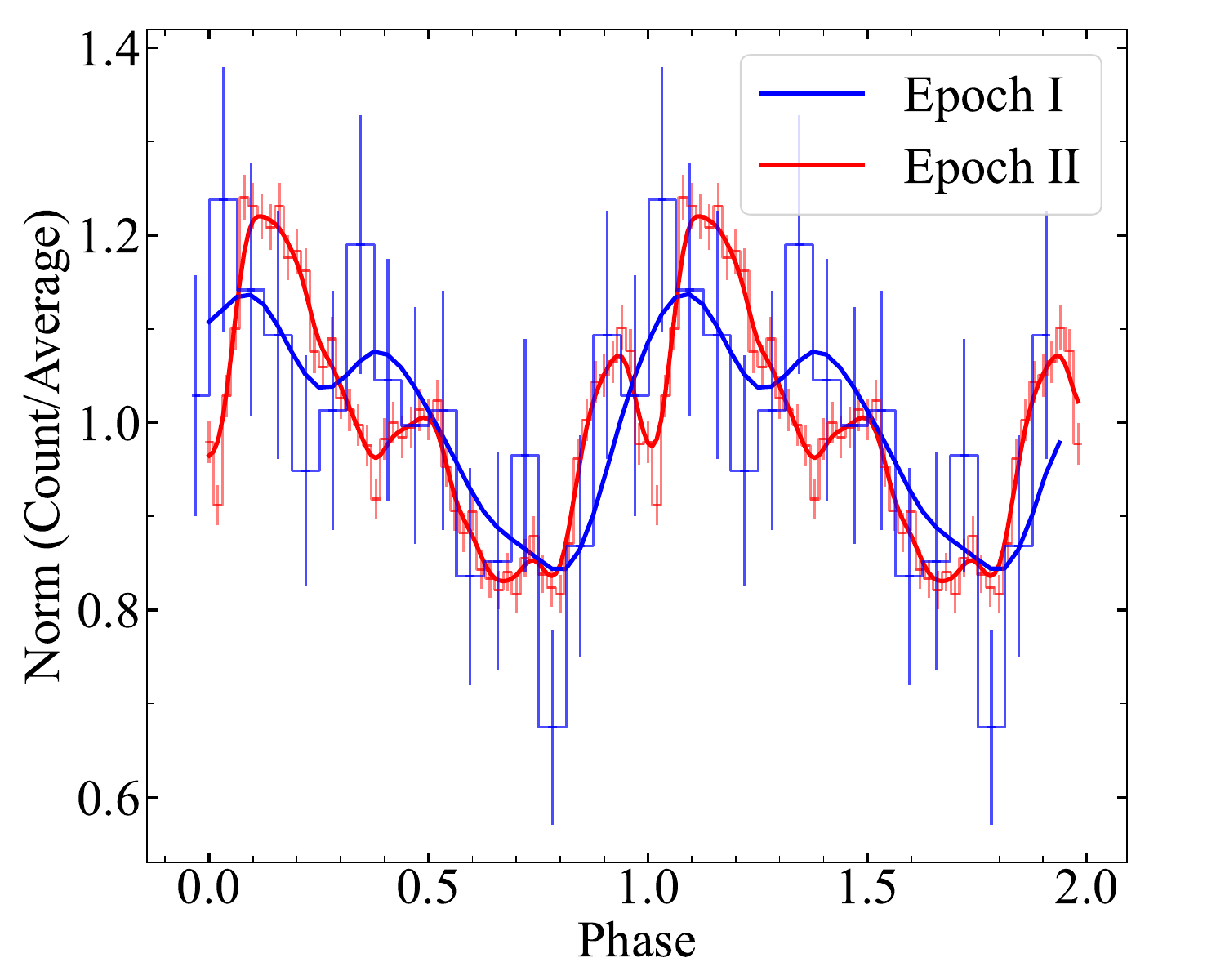}
    \caption{The average pulse profiles of 1E 1841--045 during two epochs observed by FXT in the 0.5–10 keV band. 
    Epoch I represents the pulse profile before the outburst, corresponding to Obs. ID 11908432129 on MJD 60505.57.
    Epoch II represents the average pulse profile after the outburst, spanning from MJD 60544.74 to MJD 60582.45.
    The phase of epoch I is aligned to epoch II using circular cross-correlation.
    The solid lines represent the smoothed pulse profiles after Gaussian filtering.}
    \label{fig:pro-part}
\end{figure}

\begin{figure}
    \centering  
    \includegraphics[width=\columnwidth]{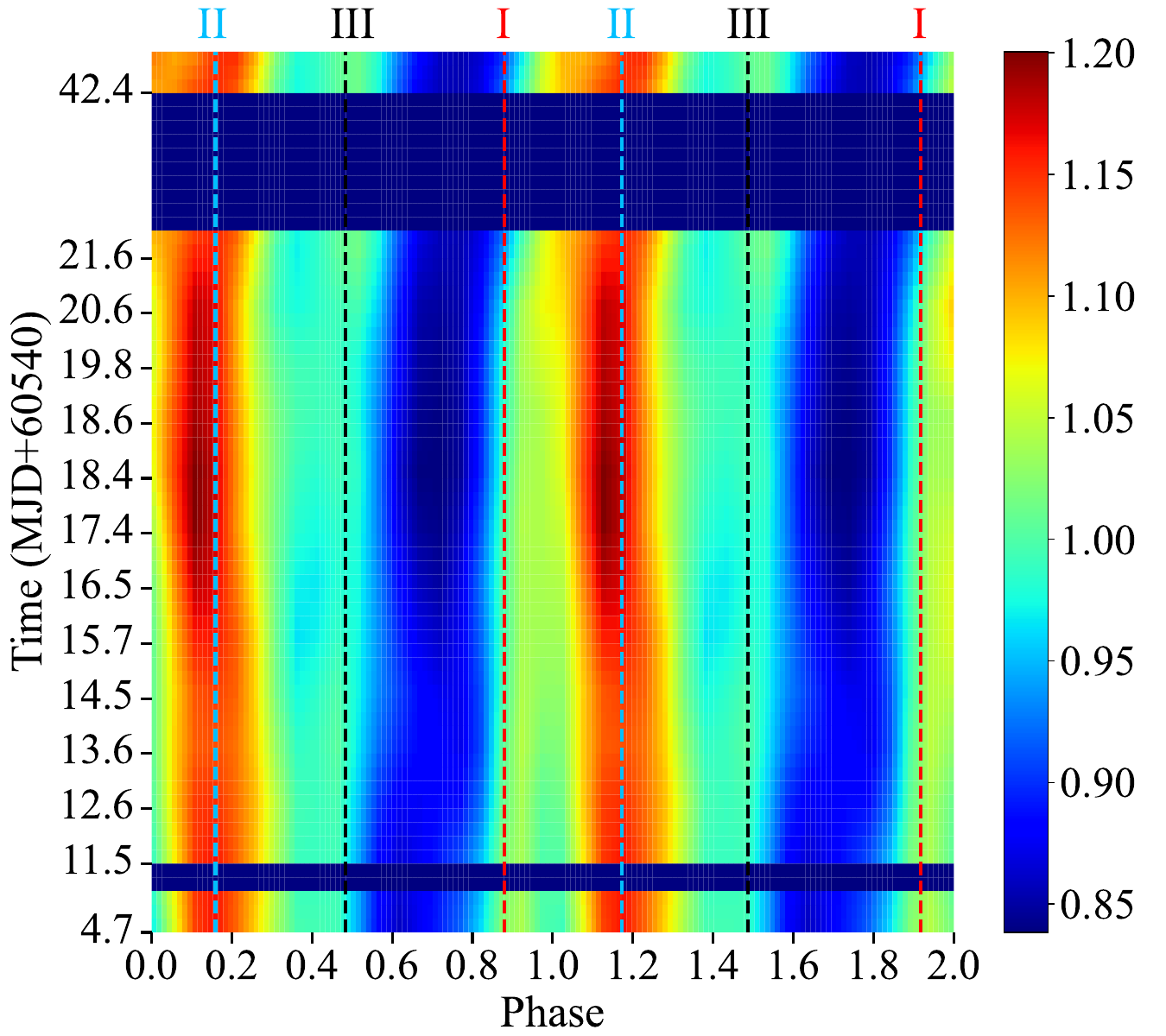}
    \caption{The two-dimensional (2D) map illustrates the evolution of the pulse profiles over time for FXT (0.5--10 keV) from MJD 60544.74 to MJD 60582.45 (Epoch II in Figure \ref{fig:pro-part}).
    The colors represent the normalized values of the pulse profile, where normalization is based on the Pulse/Average count rate.
    To create the map, 20 bins within a phase are used to fold the pulse profiles.
    For clarity, the plot is further smoothed through interpolation and Gaussian filtering.}
    \label{fig:2D}
\end{figure}

\subsubsection{Pulse profiles}\label{sec:Temporal pulse profile}
The average pulse profiles before and after the outburst in the 0.5--10 keV band, compared in Figure \ref{fig:pro-part}, show significant variability.
Epoch I represents the pulse profile before the outburst, associated with Obs. ID 11908432129 on MJD 60505.57.
Epoch II represents the average pulse profile following the outburst, spanning from MJD 60544.74 to MJD 60582.45.
The phase alignment between epoch I and epoch II is achieved by maximizing the Pearson correlation coefficient through circular shifting of the pulse profiles.
The pre-outburst profile exhibits a double-peaked structure.
Following the outburst, this is replaced by a multi-peaked structure.

The FXT pulse profiles during Epoch II display a multi-peaked structure within the 0.5–10 keV energy range, featuring peak I at approximately phase 0.9, peak II at phase 0.2, and peak III at phase 0.5, as shown in Figure \ref{fig:2D}.
The colors representing the values of the pulse profile are normalized by the Pulse/Average count rate, the red represents pulse-on phase, and the blue represents pulse-off phase.
The phase and shape of peak II (the main peak) remain relatively stable, while peaks I and III exhibit a rightward phase shift.
Over the 38-day epoch from MJD 60544 to MJD 60582, the phase-shift of peak I is $\sim 0.1$ cycle, corresponding to a rate of $\sim 2.6 \times 10^{-3}$ cycle per day, consistent with the findings reported by \citet{2025arXiv250220079Y}.
However, the significant fluctuations in the phase of peak I do not support obtaining a robust quantification through linear fitting.
Peak III appears to follow a similar trend as peak I, but its lower intensity causes it to intermittently appear and disappear.
Thus, the apparent broadening of peak III to the right in Figure \ref{fig:2D} is more likely to be caused by a reduced signal-to-noise ratio rather than a definitive phase shift.

\subsubsection{Energy-resolved pulse profile}\label{sec:Energy-resolved pulse profile}
\begin{figure}
    \centering  
    \includegraphics[width=\columnwidth]{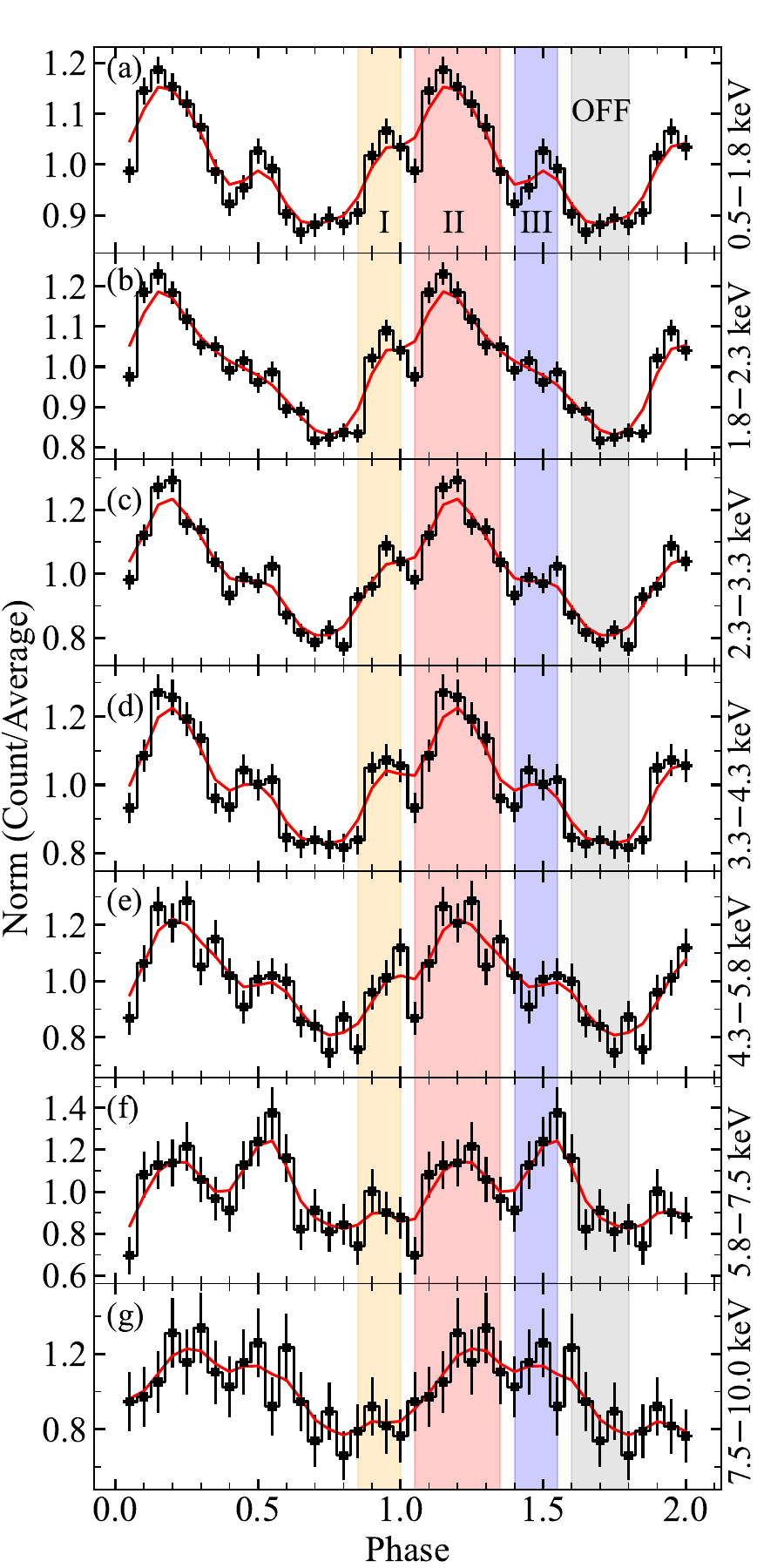}
    \caption{The evolution of the pulse profiles over energy for FXT(0.5--10 keV).
    The values of the pulse profiles are normalized by the Pulse/Average count rate, generated from the combined data spanning from MJD 60544.74 to MJD 60582.45 (Epoch II in Figure \ref{fig:pro-part}).
    The seven panels, from top to bottom, correspond to the energy bands of 0.5--1.8 keV, 1.8--2.3 keV, 2.3--3.3 keV, 3.3--4.3 keV, 4.3--5.8 keV, 5.8--7.5 keV, and 7.5--10 keV, respectively.
    The red lines in each panel represent the smoothed pulse profiles after Gaussian filtering.
    The orange, red, blue, and gray shaded areas correspond to the phase intervals of peak I, II, III, and the pulse-off period, respectively. }
    \label{fig:pro-energy}
\end{figure}

The pulse profile demonstrates a variability across different energy bands.
As illustrated in Figure \ref{fig:pro-energy}, the full energy range of the FXT (0.5--10 keV) is segmented into seven intervals.
To ensure sufficient counting statistics in each energy interval, from the top to the bottom panels, the seven energy bands correspond to 0.5--1.8 keV, 1.8--2.3 keV, 2.3--3.3 keV, 3.3--4.3 keV, 4.3--5.8 keV, 5.8--7.5 keV, and 7.5--10 keV, respectively.

The triple-peaked phase structure remains relatively stable across the 0.5--10 keV energy range, with no evidence for significant phase change. 
The normalized intensities of peaks I and II remain approximately constant at $\sim 1.1$ and $\sim 1.2$, respectively.

Peak III, however, displays distinct behavior: it persists as a minor peak with modest fluctuations below 5.8 keV, though it nearly vanishes transiently at 1.8--2.3 keV. 
Above 5.8 keV, it undergoes marked enhancement, exhibiting a clear tendency to exceed the intensity of the main peak.

\subsubsection{Pulsed fraction}\label{sec:Pulse fraction}
We further investigated the evolution of the pulsed fraction (PF) as a function of both time and energy.
After carefully accounting for and subtracting the contribution from the surrounding SNR, we calculated the PF.
Specifically, to isolate the magnetar’s emission, we folded the SNR data using the magnetar's timing ephemeris to produce a background profile.
Subsequently, after scaling the SNR background counts to account for the different size of extraction areas\footnote{%
$C_{\text{SNR, scaled}} = C_{\text{SNR, extracted}} \times \frac{A_{\text{src}}}{A_{\text{SNR}}}$,
where $C_{\text{SNR, extracted}}$ is the background counts from the SNR region, and $A_{\text{src}}$ ($A_{\text{SNR}}$) is the area of the source (background) extraction region.}, we subtracted the background pulse profile from the magnetar's.
This enabled us to isolate the magnetar's intrinsic pulse profile and, consequently, obtain its intrinsic PF from the following expression:
\begin{equation}
PF = \frac{I_{\text{max}} - I_{\text{min}}}{I_{\text{max}} + I_{\text{min}}}
\end{equation}
where $I_{\text{max}}$ and $I_{\text{min}}$ represent the maximum and minimum intensities of the intrinsic pulse profile, respectively. 

\begin{figure}
    \centering  
    \includegraphics[width=\columnwidth]{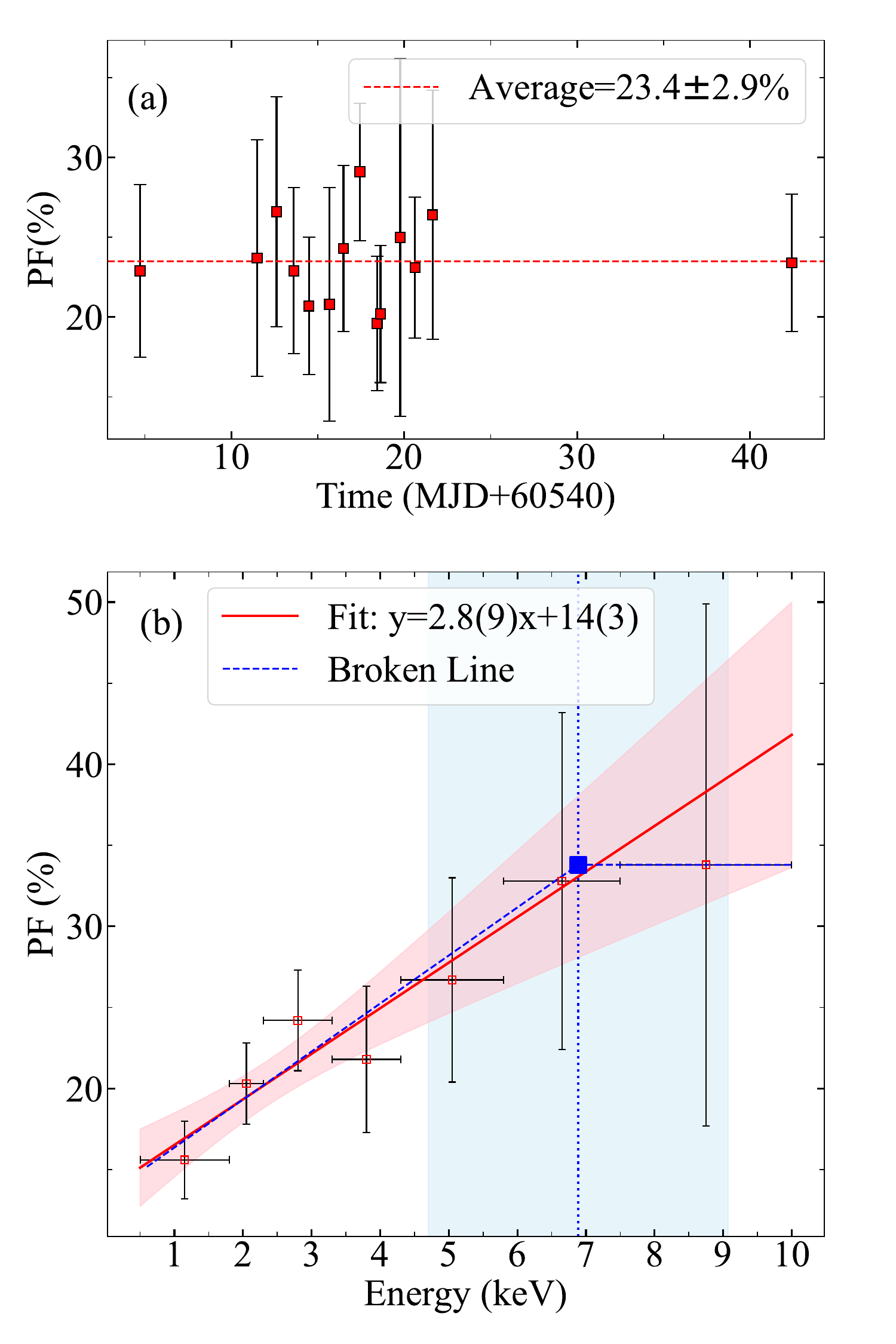}
    \caption{The evolution of the pulsed fraction (PF) of 1E 1841--045.
(a): The evolution of PF over time, with the red dashed line representing the average value.
(b): The evolution of PF over energy. 
The horizontal error bars denote the energy bin widths, with each point plotted at the bin centroid.
The red solid line shows the weighted linear fit, with the red shaded area representing the $1 \sigma$ statistical uncertainties. 
The blue dashed line shows the broken-line model fit. 
The blue vertical dotted line and the shaded region mark the turning point at $x_0 = 6.9 \pm 2.2$ keV and its $1\sigma$ uncertainty.}
    \label{fig:PF-evolution}
\end{figure}

As shown in Figure \ref{fig:PF-evolution} (a), the temporal evolution analysis reveals that the PF remains remarkably stable at $23.4\pm2.9\%$ throughout the observation period, showing no significant variations. 
The energy-dependent behavior of the PF is shown in Figure \ref{fig:PF-evolution} (b).
We first performed a weighted linear fit: $PF(\%)=2.8(9)\times E (\rm keV)+14(3)$, with $\chi^2/\nu= 1.98/5$.
The red shaded area represents the $1\sigma$ confidence interval.
The uncertainty of the weighted regression was quantified using Gaussian error propagation.
The slope of the fit remains statistically significant ($t(5)=3.1$, $p<0.05$), indicating a positive trend across the entire energy range. 
However, the widening of the confidence interval above 5.8 keV suggests that this overall trend is predominantly constrained by the low-energy points, while the high-energy behavior remains poorly constrained.
We further performed a broken-line fit, which yields a turning point at $x_0 = 6.9 \pm 2.2$ keV with $y_0 = 33.8 \pm 4.2 \%$: 
\begin{equation}
PF(\%) = 
\left\{
\begin{array}{ll}
14.1 + 2.9(1.5) \times E\ (\mathrm{keV}), & E < 6.9\ \mathrm{keV} \\[0.5em]
33.8 + 0.0(0.7) \times E\ (\mathrm{keV}), & E \ge 6.9\ \mathrm{keV}
\end{array}
\right.
\end{equation}
where the values in parentheses represent the 1$\sigma$ uncertainties of the fitted parameters.
The positive slope in the first segment ($E < 6.9$ keV) is marginally significant, while the slope in the second segment ($E \ge 6.9$ keV) is consistent with zero, indicating a plateau after the turning point. 
The fit yields $\chi^2/\nu= 1.89/3$, confirming a good description of the data.
The F-test shows no significant improvement of the broken-line model over the linear model ($P = 0.93 > 0.05$).

\subsection{Spectral analysis}\label{sec:Results-2}
We extracted the source and background spectra following the region selection criteria detailed in Section \ref{sec:Observations and Data Reduction} (Figure \ref{fig:image}).
The spectra were grouped to have at least 5 counts per bin for the fit to enable a reliable C-statistic.
The goodness of the fit was subsequently evaluated using the $\chi^2$ statistic.

\begin{figure*}
    \centering  
    \includegraphics[width=\textwidth]{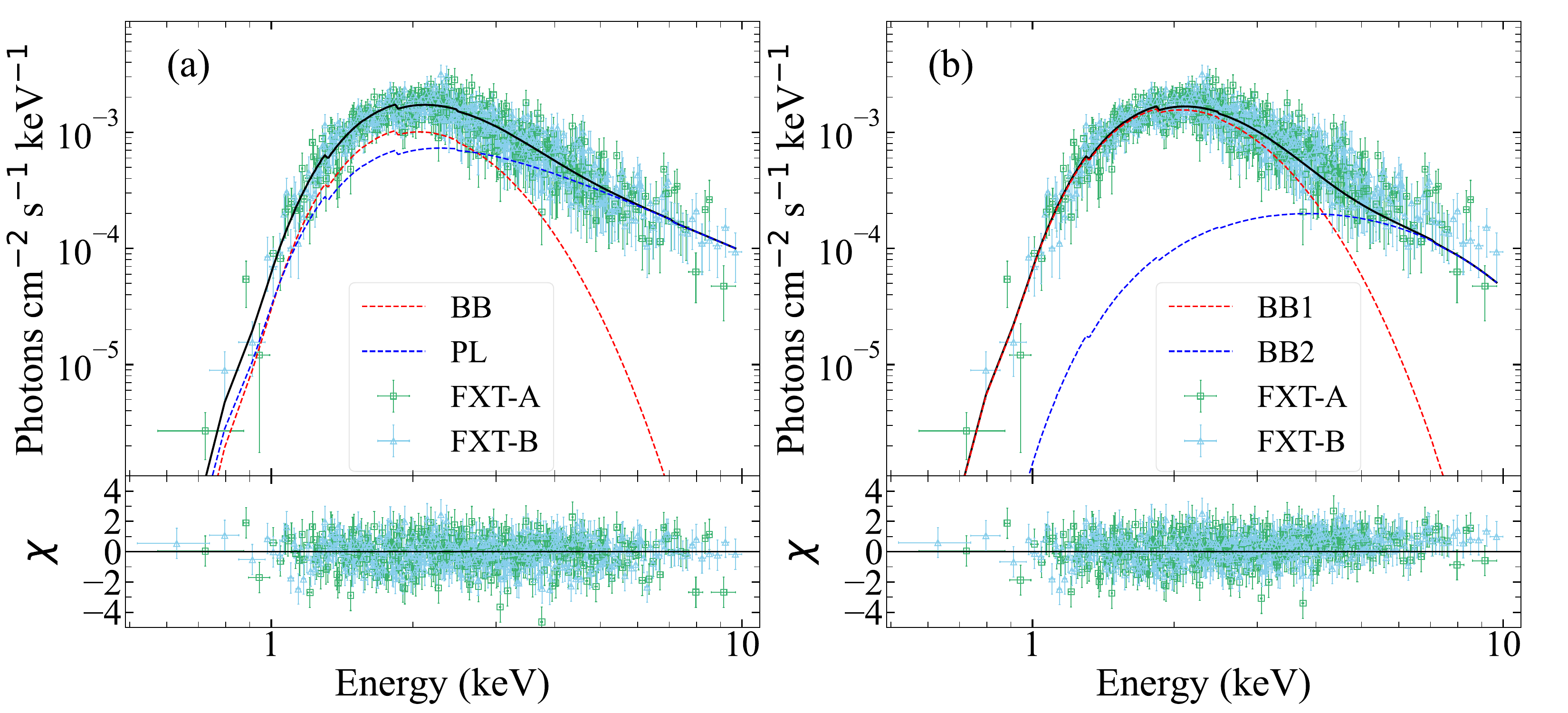}
    \caption{Spectra and residuals from Obs. ID 06800000059.
    The spectra from FXT-A and FXT-B are extracted separately and fitted jointly. 
    (a): The total model (black solid line) is shown alongside its individual components: a blackbody (red dashed line) and a power-law (blue dashed line). 
    (b): The total model (black solid line) is shown alongside its individual components: a cooler blackbody (red dashed line) and a hotter blackbody (blue dashed line).
    For clarity, the spectra have been rebinned, and the error bars represent $1 \sigma$ statistical uncertainties.}
    \label{fig:spec}
\end{figure*}

\begin{table*}
   \setcounter{table}{3}
    \begin{threeparttable}
        \caption{EP/FXT spectral fitting results from all observations of 1E 1841--045.\label{tab:spec_table2}}
        \begin{tabular*}{\textwidth}{l@{\extracolsep{\fill}}cccccccc}
            \hline  
            \hline 
            &$N_{\rm H}$&$kT_{\rm cool}$& $R_{\rm cool}$&$\rm \Gamma_{SPL}$&$kT_{\rm hot}$& $R_{\rm hot}$&$\rm \Gamma_{HPL}$&$\chi^2/ \rm dof$\\
            Model & ($\rm 10^{22}\,cm^{-2}$)& (keV) & (km) & &(keV)&(km)& &\\
            \hline 
            BB+PL	&$2.08\pm0.06$&$0.50 \pm 0.01$&$6.08\pm0.12$&$1.91 \pm 0.13$&\dots&\dots&\dots&$754/691$\\
            BB+BB	&$1.73\pm0.03$&$0.57 \pm 0.01$&$6.07\pm0.08$&\dots&$2.21 \pm 0.08$&$0.36\pm0.06$&\dots&$803/691$\\
            BB+PL+PL&$2.08^f$&$0.55 \pm 0.01$&$5.61\pm0.37$&$3.08\pm0.62$&\dots&\dots&$1.1^f$&$752/691$\\
            BB+BB+PL&$1.73^f$&$0.56 \pm 0.01$&$6.16\pm0.21$&\dots&$1.81 \pm 0.24$&$0.27\pm0.09$ &$1.3^f$&$797/691$\\
            \hline	
        \end{tabular*}
        \begin{tablenotes}[flushleft]
            \item NOTES--The source radius in kilometers is calculated from the distance of 8.5 kpc \citep{2008ApJ...677..292T}.
            The parameters marked with superscript $^f$ are fixed in the fit.
        \end{tablenotes}
    \end{threeparttable}
\end{table*}

As shown in Figure \ref{fig:spec} (a), the spectra of FXT-A and FXT-B were jointly modeled well using an absorbed blackbody plus power-law model\footnote{BB+PL model: \texttt{tbabs*(bbodyrad+powerlaw)}}.
In the joint spectral fits, the cross-normalization constant between FXT-A and FXT-B was fixed at 1, as their calibrations are known to be consistent across the employed energy band.
The source radius in kilometers is calculated from the distance of 8.5 kpc \citep{2008ApJ...677..292T}.
According to the joint fitting of all FXT observations, the typical parameters for the absorbed BB+PL model are as follows:
the hydrogen column density ($N_{\rm H}$) is $(2.08\pm0.06) \times \rm 10^{22}\,cm^{-2}$, the BB temperature ($kT_{\rm BB}$) is $0.50 \pm 0.01$ keV, the BB radius is $6.08\pm0.12$ km, and the PL photon index is $1.91 \pm 0.13$. 
The fit yields $\chi^2/ \rm dof=754/691\approx1.09$, indicating a good description of the data.
In the spectral evolution and phase-resolved analysis, we fix $N_{\rm H}$ to $2.08 \times \rm 10^{22}\, cm^{-2}$ to reduce the impact of statistical fluctuations.
The fitting results demonstrate that with degrees of freedom ranging from 296 to 794, the $\chi^2$/dof for all observations ranged from 0.98 to 1.09.

Additionally, the double blackbody model\footnote{BB+BB model: \texttt{tbabs*(bbodyrad+bbodyrad)}} 
is also applicable to the spectra of the source, as shown in Figure \ref{fig:spec} (b).
The typical parameters for the absorbed BB+BB model are as follows:
the $N_{\rm H}$ is  $(1.73\pm0.03) \times \rm 10^{22}\,cm^{-2}$, 
the cooler BB temperature ($kT_{\rm cool}$) is $0.57 \pm 0.01$ keV,
the cooler BB radius is $6.07\pm0.08$ km,
the hotter BB temperature ($kT_{\rm hot}$) is $2.21 \pm 0.08$ keV,
and the hotter BB radius is $0.36\pm0.06$ km.
The fit yields $\chi^2/ \rm dof=803/691\approx1.16$, indicating an acceptable description of the data.
Compared to the BB+BB model, the BB+PL model provides a marginally better fit (an improvement of about 6\% in $\chi^2$).
In the phase-resolved analysis, we fix $N_{\rm H}$ of BB+BB model to $1.73 \times \rm 10^{22}\, cm^{-2}$ to reduce the impact of statistical fluctuations.

We also attempted to include an additional PL component to investigate the impact of the hard PL on the soft PL (or hotter BB) component.
To do so, we fitted the averaged spectra using both a BB+PL+PL and a BB+BB+PL model.
The parameters of the hard PL component proved extremely difficult to constrain.
Even with the $N_{\rm H}$ and $kT_{\rm BB}$ fixed, the fitting only provided upper limits of 8.5 and 9.2 for the photon index of the hard PL in the two models, respectively. 
The F-test yielded p-values of 0.17 and 0.075 ($\rm P>0.05$) for the two cases, respectively, indicating that the inclusion of the hard PL component did not significantly improve the goodness of fit.
To further investigate, we fixed the hard PL photon index to the values reported by \citealt{2025ApJ...985L..34R} ($\rm \Gamma_{HPL} = 1.1$ for BB+PL+PL and $\rm \Gamma_{HPL} = 1.3$ for BB+BB+PL). 
In the BB+PL+PL model, the 6$-$10 keV absorbed flux of the soft PL is $\sim 1.54 \times 10^{-11}$ erg s$^{-1}$ cm$^{-2}$, while the hard PL contributes only $\sim 2.87 \times 10^{-13}$ erg s$^{-1}$ cm$^{-2}$, suggesting that its contribution is negligible. 
In the BB+BB+PL model, however, the hard PL contributes $\sim 5.56 \times 10^{-12}$ erg s$^{-1}$ cm$^{-2}$, compared to $\sim 1.01 \times 10^{-11}$ erg s$^{-1}$ cm$^{-2}$ from the hotter BB, indicating a more substantial contribution that affects the parameters of the hotter BB.
The complete fitting parameters are shown in Table \ref{tab:spec_table2}. 

\subsubsection{Phase-averaged spectral evolution}\label{sec:Spectral Evolution}
\begin{figure}
    \centering  
    \includegraphics[width=\columnwidth]{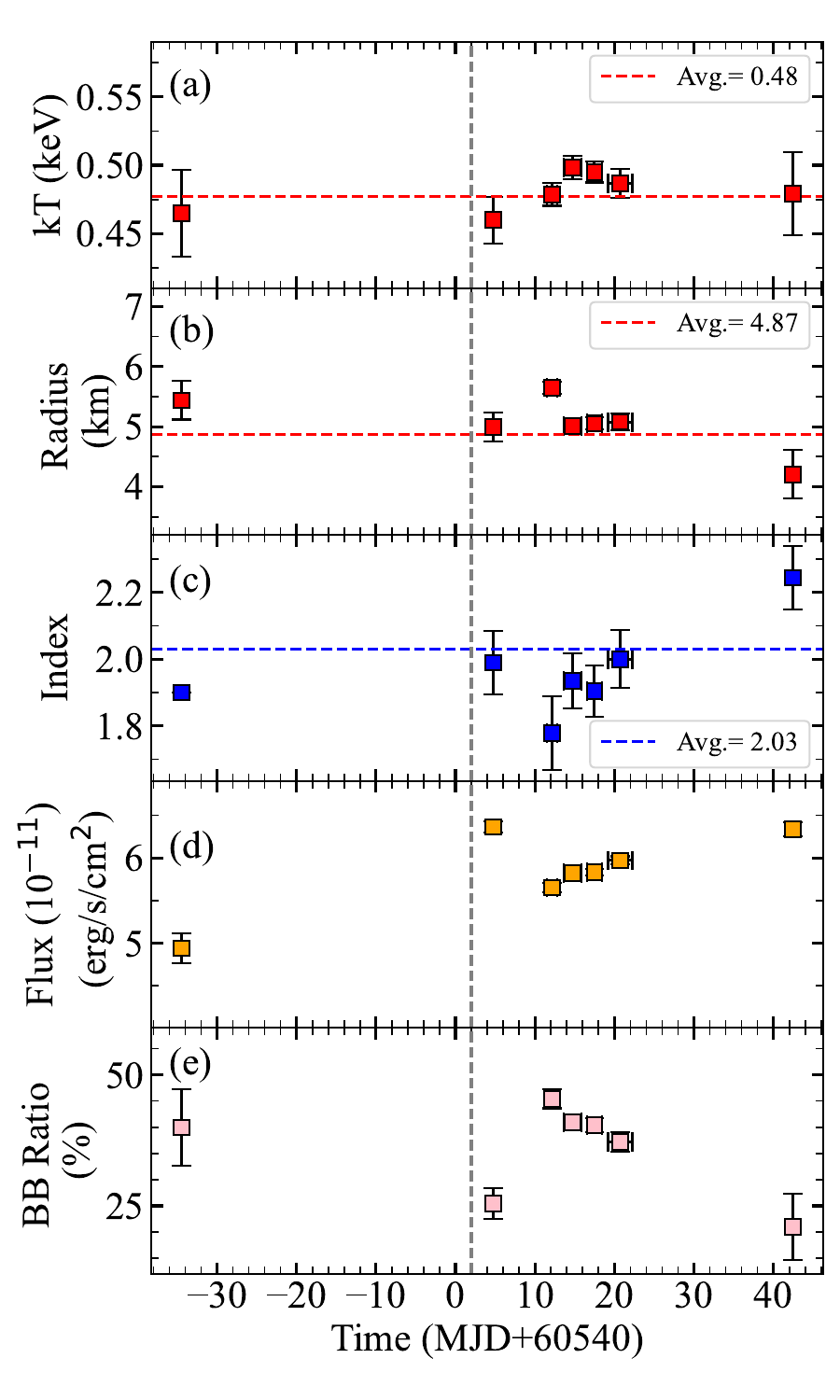}
    \caption{Temporal evolution of spectral parameters: blackbody temperature, emission radius, power-law photon index, unabsorbed total flux, and the fractional contribution of the blackbody flux to the total flux.
    The source radius in kilometers is calculated from the distance of 8.5 kpc \citep{2008ApJ...677..292T}.
    The dashed horizontal line indicates the mean value across all parameters.
    The dashed vertical line indicates X=2, marking the onset of the outburst.
    The errors are calculated with $1 \sigma$ level uncertainties.}
    \label{fig:spectralevo}
\end{figure}

We perform the spectral fitting for each observation using the BB+PL model, and the results are shown in Table \ref{tab:spec_table}.
In Figure \ref{fig:spectralevo}, the evolution of the BB temperature ($kT_{\rm BB}$), BB radius, photon index of PL, unabsorbed flux in 0.5--10 keV, and proportion of BB component to the total flux with time are displayed from top to bottom panels.

\begin{figure*}
    \centering  
    \includegraphics[width=\textwidth]{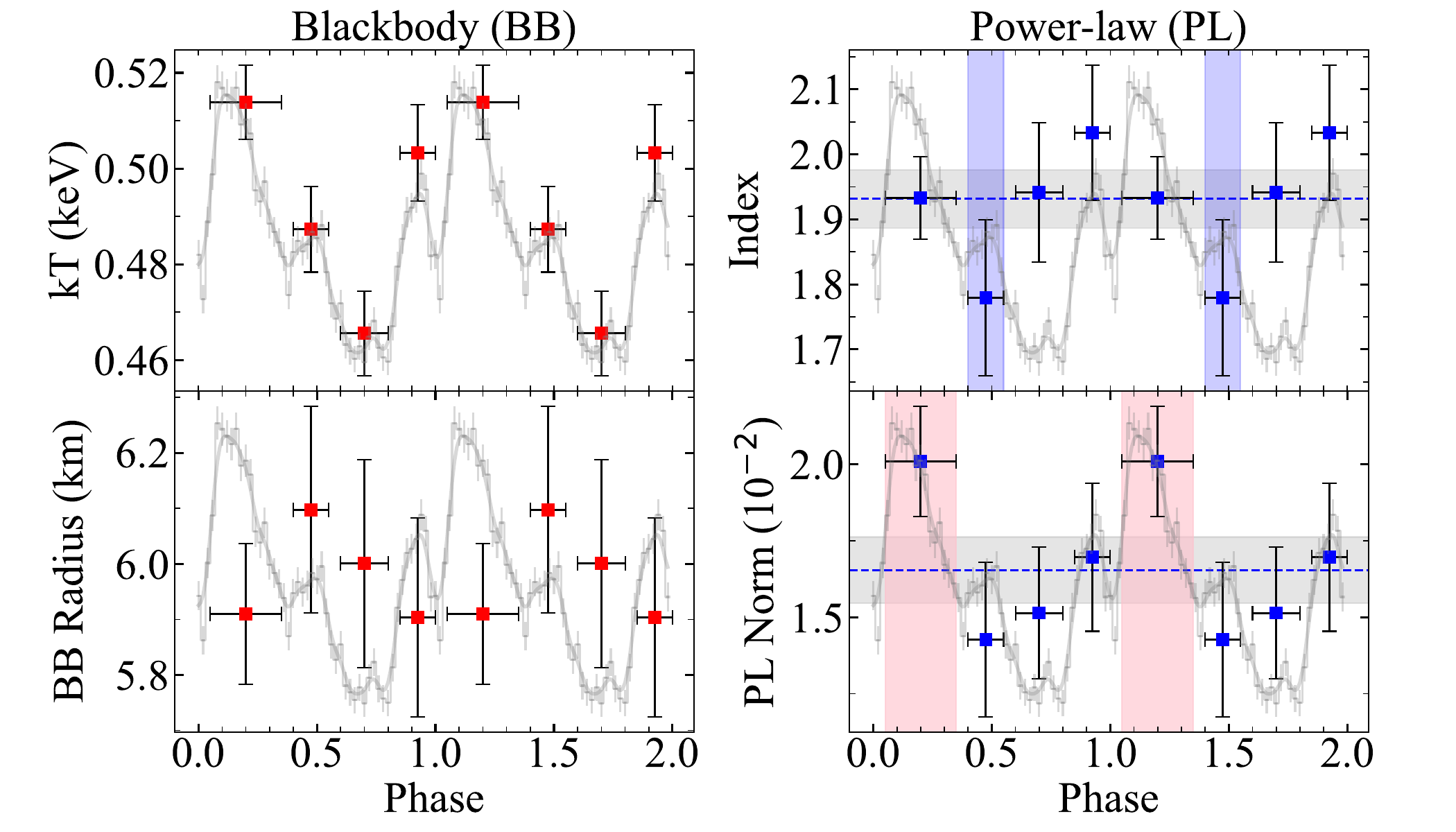}
    \caption{Fitting results of the phase-resolved spectral analysis of the BB+PL model.
    Solid lines represent the pulse profiles as shown in Figure \ref{fig:pro-part} (Epoch II).   
    The blue dashed horizontal lines represent the constant fits for the photon index ($\Gamma = 1.93 \pm 0.04$) and the PL normalization ($0.017 \pm 0.001$), with gray shaded $1\sigma$ uncertainties. 
    The blue and pink vertical bands mark peak III and the main peak (peak II), respectively.
    The source radius in kilometers is calculated from the distance of 8.5 kpc \citep{2008ApJ...677..292T}.
    The errors are calculated with $1 \sigma$ level uncertainties.} 
    \label{fig:phase_resolved}
\end{figure*}

As observed in previous outbursts, the source showed no detectable signs of temporal variation in spectral parameters. 
In Figure \ref{fig:spectralevo} (a) and (b), the BB remains approximately stable at $kT_{\rm BB}$ $\sim$ 0.48 keV temperature, with an average emitting radius of 4.87 km. 
This source radius in kilometers was derived based on the distance of 8.5 kpc \citep{2008ApJ...677..292T}.
Conversely, when employing a distance of 5.8 kpc \citep{2018AJ....155..204R},  the average emitting radius is determined to be 3.32 km.
It is important to note that the 8.5 kpc distance has been used consistently in prior studies. 
To ensure comparability with these earlier works \citep{2025ApJ...985L..34R, 2025ApJ...985L..35S, 2025arXiv250220079Y}, we have opted to continue using the 8.5 kpc distance throughout this analysis.
Figure \ref{fig:spectralevo} (c) reveals a mean photon index of 2.03; although the last data point exhibits a slight upward trend, the large associated uncertainty suggests this deviation is likely due to statistical fluctuations.
Due to the limited observation time, the first data point's spectrum lacks sufficient counting statistics, making it unable to effectively constrain the error of photon index. 
However, a BB component alone cannot adequately describe the spectrum, so the PL component is still required.
Consequently, the photon index of first point is set at a fixed value of 1.9.
The evolution of unabsorbed flux, displayed in Figure \ref{fig:spectralevo} (d), shows an increase from a pre-outburst level of $4.94\times10^{-11} \rm erg\, s^{-1} cm^{-2}$ to a post-outburst value of $\sim 6 \times10^{-11} \rm erg\, s^{-1} cm^{-2}$, representing a $\sim 20\%$ rise. 
However, systematic uncertainties from the limited pre-outburst sampling (single observation) may affect this estimate.
Figure \ref{fig:spectralevo} (e) presents the BB flux ratio, defined as the unabsorbed BB flux divided by the total unabsorbed flux.
The BB ratio exhibits a decrease at the first observation after the outburst.

\subsubsection{Phase-resolved spectral analysis}\label{sec:Phase-resolved Spectral Analysis}
\begin{figure*}
    \centering  
    \includegraphics[width=\textwidth]{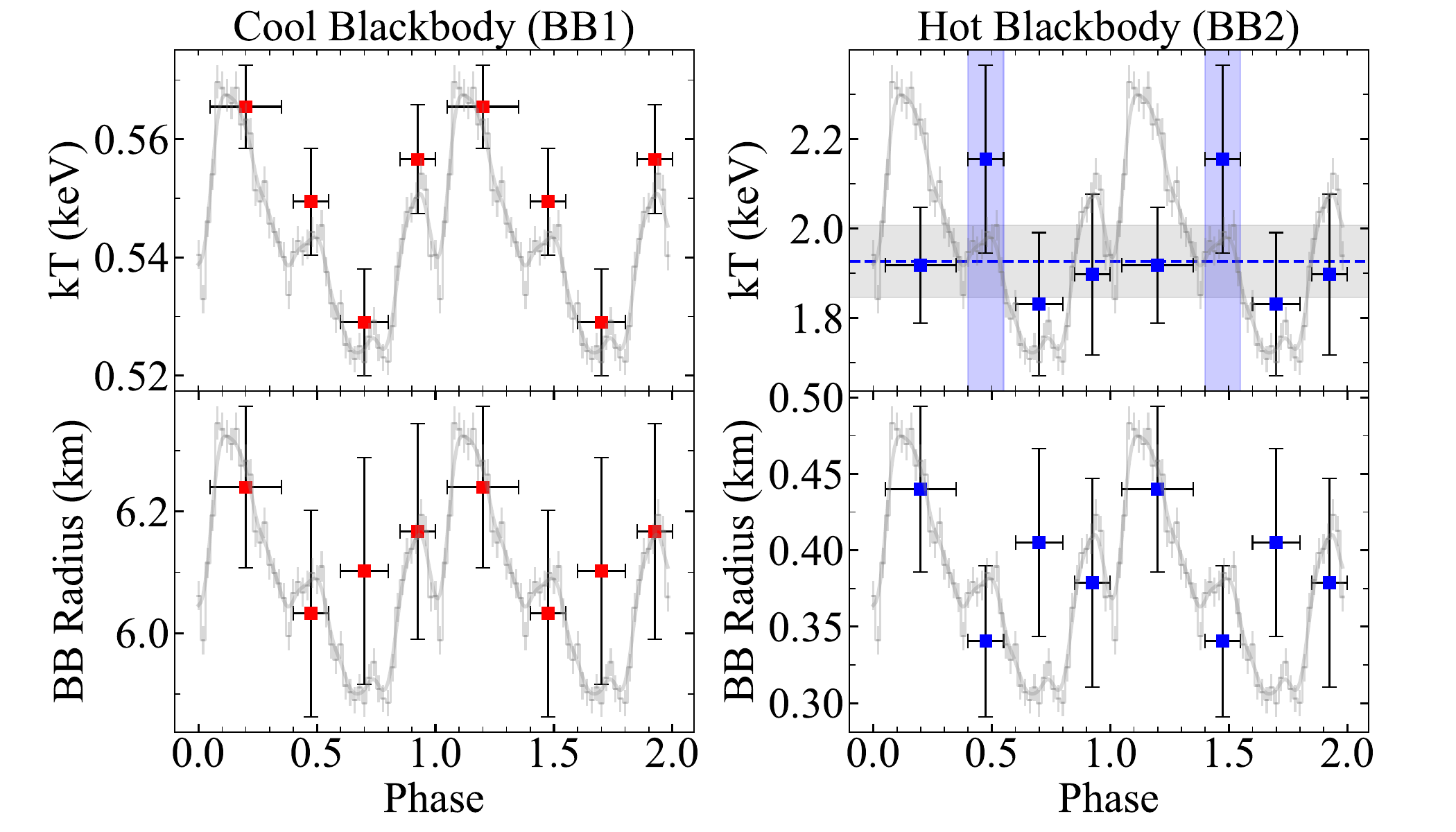}
    \caption{Fitting results of the phase-resolved spectral analysis of the BB+BB model.
    The blue dashed horizontal line indicates the constant fit result for the hotter BB temperature ($1.93 \pm 0.08$ keV), with the gray shaded region representing the $1\sigma$ uncertainty.
    The blue vertical band highlights the phase interval corresponding to peak III.
    The source radius in kilometers is calculated from the distance of 8.5 kpc \citep{2008ApJ...677..292T}.
    The errors are calculated with $1 \sigma$ level uncertainties. }
    \label{fig:phase_resolved2BB}
\end{figure*}

In order to analyze the possible transition of the spectral properties of each feature in the pulse profile, we study the phase-resolved spectral analysis.
The phase-resolved spectra are modeled well using a BB+PL model combination.
Figure \ref{fig:phase_resolved} presents the phase-resolved spectral fitting results. 
The gray background traces represent the pulse profile, while the red and blue data points correspond to the BB and PL spectral components, respectively.
The horizontal error bars on the data points indicate the phase bin widths used in the analysis.
The entire cycle is divided into four phase intervals, peak I (0.85--1.00 phase), peak II (0.05--0.35 phase), peak III (0.40--0.55 phase), and the minimum pulse phase (0.60--0.80 phase).
Phase intervals with low signal-to-noise ratio between the defined peaks were excluded to ensure reliable spectral analysis.
The four phase intervals are also indicated by the shaded areas in Figure \ref{fig:pro-energy}.
The fitting yields $\chi^2$/dof ranging from 1.01 to 1.07 across all four phase intervals, with degrees of freedom ranging from 582 to 691.

Our analysis reveals a clear positive correlation between the $kT_{\rm BB}$ of BB and the pulse profile intensity. 
The BB reaches its maximum temperature value of $0.514\pm0.007$ keV during the main peak (Peak II), which corresponds to the highest intensity pulse phase. 
The temperature gradually decreases in the secondary peak (peak I), with $0.503\pm0.010$ keV. 
The minor peak (peak III) temperature is even lower at $0.487\pm0.009$ keV.
The minimum $kT_{\rm BB}$, which is found to be $0.465\pm0.008$ keV, is observed during the pulse-off phase.
In contrast to the $kT_{\rm BB}$ variations, the BB radius remains constant across all pulse phases, showing no statistically significant phase-dependent modulation.

The PL component exhibits complex variability. 
A constant fit to the four photon index points yields $\Gamma = 1.93 \pm 0.04$ with $\chi^2/\mathrm{dof} = 3.2/3$, indicating no significant variation of the photon index with phase (horizontal shaded region in Figure \ref{fig:phase_resolved}). 
The value at peak III ($\Gamma = 1.78 \pm 0.12$) is slightly lower than the constant level, but the deviation is not statistically significant ($\Delta \Gamma = 0.15 \pm 0.13$, corresponding to $1.15 \sigma$). 
The energy-resolved pulse profile analysis (Figure \ref{fig:pro-energy}) shows that peak III is more prominent at higher energies, which is potentially consistent with a harder spectrum, though the photon index variation itself is not statistically significant. 
The normalization parameter of the PL component exhibits a potential positive correlation with the pulse profile, reaching its maximum value during the main peak.
Fitting the four data points with a constant model yields PL $\rm Norm=0.017 \pm 0.001$ 
 with a $\chi^2$/dof = 6.28/3 ($\rm P = 0.099>0.05$), indicating that the data do not significantly deviate from a constant. 
The Peak II point (red area) is marginally elevated, with $\Delta \rm Norm = 0.0036 \pm 0.0021$ above the constant level, corresponding to a 1.71 $\sigma$ deviation.

For comparison, we also performed phase-resolved spectral fitting using a BB+BB model, as shown in Figure \ref{fig:phase_resolved2BB}.
The fit yields the $\chi^2$/dof for all phase intervals ranging from 1.02 to 1.18, with degrees of freedom ranging from 582 to 691.
The cooler BB temperature $kT_{\rm cool}$ exhibits a clear positive correlation with the pulse profile intensity, similar to the behavior observed for the single BB component in the BB+PL model.
The radius of the cooler BB remains constant across all phases, showing no significant modulation.

The hotter BB component shows a different behavior. 
A constant fit to the four $kT_{\rm hot}$ points yields $1.93 \pm 0.08$ keV with $\chi^2/\mathrm{dof} = 2.6/3$, indicating no significant variation with phase (horizontal shaded region in Figure \ref{fig:phase_resolved2BB}).
The value at peak III slightly increases, though the deviation is not statistically significant ($\Delta kT_{\rm hot} = 0.24 \pm 0.19$ keV, 1.26 $\sigma$). 
This behavior is similar to that of the photon index in the BB+PL model, where peak III also showed a marginal hardening.
The radius of the hotter BB remains constant within uncertainties across all phases.

\subsection{Radio results}\label{sec:Results-4}
Data processing for FAST in the radio band is conducted on FAST cluster servers, with the main search program based on the single-pulse search module in PRESTO. 
Based on observations from the Parkes telescope, we set 529 as the reference dispersion measure (DM) value for data analysis of this source and established a search range of 400--600 (step size is 0.04) for DM.
Preliminary search results were filtered using a signal-to-noise ratio threshold of 6.
The filtered results indicate no significant signal from 1E 1841--045 within the DM range of 400--600 at 1--1.5 GHz.
The upper limit for single-pulse flux density is 2.6 mJy across the range of pulse timescales 0.05--50 ms, which is consistent with the upper limits derived for magnetars from \citet{2025ApJ...979..122B} and \citet{2020Natur.587...63L}.

\section{Discussion} \label{sec:Discussion}
We have analyzed the EP observations of 1E 1841--045 during its 2024 active period and performed the timing and spectral analysis of the X-ray persistent emission. The main results are as follows:

\textbullet\, During the decay of the outburst, we detected a stable spin period of 0.084699564(6) Hz, with a derivative of $-3.04(5) \times 10^{-13}$ Hz/s, as shown in Table \ref{tab:tabletiming}.

\textbullet\, The pulse profile exhibited a multi-peaked structure. 
The secondary peak (peak I) underwent a phase shift of $\sim$ 0.1 cycle, while the minor peak (peak III) appeared and disappeared intermittently.
The energy-resolved pulse profile revealed a stable triple-peaked structure, with the minor peak becoming significantly enhanced above 5.8 keV, as illustrated in Figures \ref{fig:2D} and \ref{fig:pro-energy}.
Moreover, the pulsed fraction of the integrated pulse profile increased linearly with energy, as depicted in Figure \ref{fig:PF-evolution}. 

\textbullet\, The phase-averaged spectra were well-fitted by an absorbed blackbody plus power-law model, showing a $\sim 20\%$ flux increase in the soft (0.5--10 keV) X-ray band following the outburst, as shown in Figures \ref{fig:spec} and \ref{fig:spectralevo}. 

\textbullet\, Phase-resolved analysis indicated a positive correlation between the blackbody temperature ($kT_{\rm BB}$) and the pulse profile intensity, with $kT_{\rm BB}$ tracing the pulse profile.
Additionally, there was spectral hardening of the power-law component at the minor peak, as shown in Figure \ref{fig:phase_resolved}.

\subsection{Timing and pulse profile}\label{sec:Timing and pulse profile}
The occurrence of spin-up glitches during the onset of outburst is well documented in the persistently monitored X-ray magnetars \citep{2014ApJ...784...37D}, just like the glitch observed in 1E 1841--045 at 2024 outburst monitored by NICER \citep{2025arXiv250220079Y}.
The spin-up glitches are typically explained by the superfluid mechanism in pulsars, where angular momentum is transferred from the faster-rotating superfluid component in the NS's inner crust to the slower-rotating outer crust that is being braked by magnetic dipole radiation \citep[e.g.,][]{2012puas.book.....L}.
For 1E 1841--045, the initial FXT ToO observation was conducted on MJD 60544.74, approximately 1.74 days later than the reported glitch epoch (MJD 60543; \citealt{2025arXiv250220079Y}). 
Due to the limited pre-glitch coverage consisting of only one brief monitoring observation, FXT's observational gap prevented the detection of this glitch event.
This underscores the need for continuous monitoring during outburst onset.

Before and after the outburst, there was a significant change in the pulse profile shape, as shown in Figure \ref{fig:pro-part}. 
The pre-outburst bimodal structure evolved into a triple-peaked profile, with the strongest main peak splitting into two components and the newly emerged secondary peak (peak I) displaying a phase width of approximately 0.15.
The smallest minor peak (peak III) exhibited no detectable variation.
This pulse profile evolution was also independently detected by NICER observations \citep{2025arXiv250220079Y}.
Following the outburst, FXT similarly detected a phase shift in the newly emerged peak, which exhibited gradual rightward drift toward the main peak, suggesting an eventual merging trend. 
FXT's monitoring duration was substantially shorter than NICER's and insufficient for deriving quantitative drift measurements. 
However, we obtained a crude estimate of $\sim 2.6 \times 10^{-3}$ cycle per day.
This value exceeds NICER's precise measurement of $ 7.3(5) \times 10^{-4}$ cycle per day \citep{2025arXiv250220079Y}, yet we contend that FXT independently corroborates NICER's findings.

The pulse profile also exhibits significant variability across different energy bands.
Historical records indicate that 1E 1841--045 shows a flatter pulse profile in hard X-rays \citep[e.g.,][]{2004ApJ...613.1173K, 2013ApJ...779..163A}.
RXTE observations revealed a transition in the main pulse structure when energy exceeded $\sim$ 9 keV, with the dominant peak moving from phase $\sim$ 0.3 to phase $\sim$ 0.7 \citep{2004ApJ...613.1173K}.
NuSTAR detected the same transition phenomenon, though occurring at a higher energy boundary at $\sim$ 11 keV.
Considering the measurement uncertainties, the difference between $\sim$ 9 keV and $\sim$ 11 keV was insignificant \citep{2013ApJ...779..163A}.
During the 2024 outburst, NuSTAR observations demonstrated that while the 10--20 keV profile was dominated by a peak at phase $\sim$ 0.75, the 3--10 keV profile was dominated by a peak at phase $\sim$ 0.5.  
In contrast to historical records, the previously observed peak at phase $\sim$ 0.3 split into two narrower peaks at phases $\sim$ 0.2 and $\sim$ 0.5 \citep{2025arXiv250220079Y}.
FXT provided higher energy-resolution pulse profiles in the 0.5--10 keV soft X-ray band, clearly showing that the dominant peak transition occurred at $\sim$ 5.8 keV (see Figure \ref{fig:pro-energy}). 
Above this threshold energy, the primary peak dominance transitioned from Peak II (corresponding to phase $\sim$ 0.5 of NuSTAR mentioned earlier) to Peak III (corresponding to phase $\sim$ 0.75).
The discrepancy between $\sim$ 5.8 keV and the historically recorded $\sim$ 9 keV or $\sim$ 11 keV should not be overlooked, and it suggests potential changes in the energy boundary for pulse profile transitions. 
This energy boundary has been associated with the spectral break \citep{2013ApJ...779..163A}.
FXT's spectral analysis showed no other notable changes beyond $\sim$ 5.8 keV except for the increasing dominance of the PL component.
The phase-resolved spectra also indicate that Peak III has a potentially harder spectrum than other phase intervals.
This indicates Peak III originated from a different location than Peaks I and II.

The pulsed fraction (PF) demonstrates minimal temporal variability, as evidenced by the FXT mean PF remaining constant at $23.4\pm2.9\%$.
Both NuSTAR and NICER observations exhibit a similar steady trend, but the derived PF values are $\sim 11\%$ and $\sim 6\%$, respectively \citep{2025arXiv250220079Y}, which are lower than the PF value of FXT.
IXPE measurements reveal a PF of $21.2\pm1.5\%$ in the 2--4 keV energy band and $27.6\pm7.3\%$ in the 4--8 keV energy band.
While a slight increase in PF at higher energies is observed, this trend lacks strong statistical significance, suggesting minimal energy dependence in the IXPE energy range \citep{2025ApJ...985L..34R, 2025ApJ...985L..35S}.
In contrast, in previous observations of the source, the PF for 1E 1841--045 exhibited an energy-dependent increase across broad energy bands. 
According to \citet{2006ApJ...645..556K}, the PF reaches $\sim 25\%$ at 20 keV, increasing to nearly $100\%$ at energies above 100 keV.
\citet{2013ApJ...779..163A} also reported an increase in PF from $10\%$ at 2 keV to $20\%$ at 80 keV.
During the 2024 outburst, FXT's soft X-ray measurements show that the 
PF exhibits an increasing trend with energy below $\sim5$ keV, rising from $\sim15\%$ at $\sim0.5$ keV to $\sim 25\%$ at $\sim5$ keV, with a linear growth rate of $2.8\pm0.9\%$ per keV.
Beyond $\sim5$ keV, the behavior is poorly constrained due to large uncertainties.

\subsection{Soft X-ray spectra}\label{sec:Soft X-ray spectra}
The 0.5--79 keV X-ray spectrum of 1E 1841--045 shows some variability in the fitting results \citep[e.g.,][]{2004ApJ...613.1173K, 2010PASJ...62.1249M, 2015ApJ...815...15W}.
The previous fitting results indicate the spectrum above 3 keV is dominated by non-thermal component, and in this case, a model with two PL components can fit the 3--79 keV spectrum \citep{2017ApJS..231....8E, 2025arXiv250220079Y}.
After the outburst, the increased soft component makes the thermal component more significant in the model.
For the soft 0.5--10 keV spectrum, the BB+PL combination provides a good fit, with Swift, XMM-Newton and Chandra observations giving a typical BB temperature of $\sim$ 0.46 keV and PL photon index ranging from 1.76 to 2.07 \citep{2013ApJ...779..163A}.
The best-fit results from FXT's 0.5--10 keV spectra show the typical BB temperature is $0.50 \pm 0.01$ keV, and the PL photon index is $1.71 \pm 0.13$, which shows good agreement with the Swift measurements.
The FXT freely fitted $N_{\rm H}$ value of $(2.08\pm0.06) \times \rm 10^{22}\, cm^{-2}$ is slightly lower than Swift's fitted value of $(2.23\pm0.25) \times \rm 10^{22}\, cm^{-2}$ \citep{2013ApJ...779..163A}.
Other analyses also provided different measurement values between $\sim (2.4-2.9) \times \rm 10^{22}\, cm^{-2}$ \citep{2011ApJ...740L..16L, 2013ApJ...779..163A, 2017ApJS..231....8E, 2025ApJ...985L..34R, 2025ApJ...985L..35S}.
However, when considering factors such as instrumental differences, varying activity periods, and efficiency corrections for Kes 73's influence, this discrepancy may be negligible.

1E 1841--045 showed almost no significant changes in its spectral parameters following the outburst, with the soft X-ray flux increasing by $\sim 20\%$ compared to pre-outburst levels.
In the case of another active magnetar, SGR J1935+2154 \citep{2020ApJ...904L..21Y, 2025ApJ...980...99Y}, the decay pattern revealed that the BB and PL components did not decrease in tandem during the outburst decay. 
The BB component rapidly weakened within a few hours, while the PL component decayed slowly over several days.
Eventually, the source entered a relatively stable phase characterized by constant spectral parameters.
The persistent emission outburst of SGR J1935+2154 was more intense than that of 1E 1841--045, resulting in more discernible differences in the decay of thermal and non-thermal spectral components. 
Specifically, the BB flux ratio increased from $\sim 10\%$ during the outburst to $\sim 60\%$ in the quiescent state \citep{2025ApJ...980...99Y}.
We attempted to track the evolution of the BB and PL components in 1E 1841--045. 
Before the outburst, the BB flux fraction was $\sim 40\%$, which dropped to $\sim 25\%$ post-outburst, and then eventually recovered to $\sim 40\%$. 
However, due to sparse observational coverage, a definitive picture of the BB ratio evolution could not be established.
Nevertheless, when compared to SGR J1935+2154, the weaker outburst activity of 1E 1841--045, along with its less pronounced BB ratio evolution, appears to be in harmony.

Previous studies have discussed the emission mechanisms of 1E 1841--045's pulsed emission \citep{2013ApJ...779..163A, 2015ApJ...807...93A, 2025arXiv250220079Y}. 
Regardless of whether it is explained by the "outflow" model \citep{2009ApJ...703.1044B, 2013ApJ...762...13B} or the "plastic motions" model \citep{2002ApJ...574..332T, 2025arXiv250220079Y}, the emission is thought to be of non-thermal origin.
The soft X-ray emission originates from hot spots generated by high-energy charged particles bombarding the NS's surface, such as the footpoints of twisted magnetic field lines \citep{2016ApJ...833..261B}.
Meanwhile, the non-thermal emission in the high-energy band can extend below 10 keV \citep{2013ApJ...779..163A}. 
Thus, a single BB component cannot adequately fit the 0.5--10 keV spectrum.
Instead, an additional component is required, and either a BB+BB model (accounting for multiple thermal components) or a BB+PL model (combining thermal and non-thermal emission) can be employed to achieve a better fit.
The evolution of the low-temperature BB ratio actually reflects the heating process of the hot spots by high-energy charged particles.

The phase-resolved spectra of the soft X-ray emission do not fully agree with the results reported by \citet{2013ApJ...779..163A}.
A distinct positive correlation is observed between BB temperature and the intensity of the pulse profile. 
This marks the first detection of significant phase-dependent variations in the BB component of this source.
The photon index of the PL component remains relatively stable across different phases, except for a slightly harder spectrum observed at peak III of the pulse profile. 
This finding is further supported by the energy-resolved pulse profile analysis.
Notably, NuSTAR's phase-resolved spectroscopy reveals the same trend.
A single PL component provides an excellent fit to the hard X-ray pulse-phase spectra, and it exhibits a harder energy spectrum at the third peak \citep{2025arXiv250220079Y}. 
This implies that the PL component in the soft X-ray energy range exhibits the same behavior as the PL component in the hard X-ray energy range.
This consistency may indicate a common non-thermal origin for the pulsed emission across the two energy bands.
Therefore, from a physical standpoint, these results indicate that the BB+PL model may provide a more suitable description of the soft X-ray spectra than the BB+BB model, though the current data cannot statistically distinguish between the two models.
The normalization of the PL component is consistent with a constant within uncertainties. 
However, a slight excess is observed at Peak II ($\Delta \mathrm{Norm} = 0.0036 \pm 0.0021$, corresponding to a $1.71\sigma$ deviation), but its statistical significance is marginal. 
Given this marginal significance, deeper observations are needed to confirm whether this excess reflects a genuine non-thermal contribution to the pulsed emission.

The spatial resolution of EP offers a significant advantage over instruments like NICER, particularly for analyzing sources located within the circular ring of SNR.
Although the separability of the SNR has a relatively minor impact ($\sim 2\%$) on the calculation of the source's PF in 1E 1841--045, it still helps to mitigate any adverse effects from the surrounding region.
More crucially, the SNR exerts a substantial influence on the spectral fitting of the source. 
Therefore, separating the SNR emission in the soft X-ray band is important for accurately measuring the neutral hydrogen column density ($N_{\rm H}$) towards the source. 

\begin{acknowledgments}
This work has made use of data from the EP mission, as well as data and software provided by the Einstein Probe Science Center (EPSC). 
This work is also based on data obtained by Insight-HXMT, NICER, Swift, FAST missions.
This work is supported by the National Key R\&D Program of China (2021YFA0718504). 
The authors thank supports from the National Natural Science Foundation of China under Grants U2038103, U2038101, U2038102 and 12373051. This work is also supported by International Partnership Program of Chinese Academy of Sciences (Grant No.113111KYSB20190020).
F.C.Z. is supported by a Ram\'on y Cajal fellowship (grant agreement RYC2021-030888-I), the Spanish grant ID2023-153099NA-I00, and the program Unidad de Excelencia Maria de Maeztu CEX2020-001058-M.
We would like to thank Dr. George Younes for his helpful comments and suggestions on this project.
We also thank the anonymous referee for the insightful comments and suggestions that improved the quality of this work.
\end{acknowledgments}

%

\software{ASTROPY
\citep{2013A&A...558A..33A,2018AJ....156..123A},  
          XSPEC \citep{1996ASPC..101...17A}, and  STINGRAY \citep{2019ApJ...881...39H}.
          }





\bibliography{sample631}{}
\bibliographystyle{aasjournal}



\end{document}